# Radius Study of Ten Transiting Hot Jupiter Exoplanets with Ground-Based Observations

Davoudi F.[1]; Poro A.[1], Paki E.[2]; Mirshafie P.[2]; Ahangarani F.[2]; Farahani A.[1]; Roshana M.[2]; Abolhasani F.[2]; Zamanpour Sh.[2]; Lashgari E.[1]; Modarres S.[1]; Mohandes A.[2]

[1]The International Occultation Timing Association-Middle East section, info@iota-me.com
[2]Exoplanet Transit Project (ETP), Transit department, IOTA-ME, Iran

**Abstract**
In this research, 14 light curves of 10 hot Jupiter exoplanets available on Exoplanet Transit Database (ETD) were analyzed. We extracted the transit parameters using EXOFAST software. Finally, we compared the planet's radius parameter calculated by the EXOFAST with the value at the NASA Exoplanet Archive (NEA) using the confidence interval method. According to the results obtained from this comparison, there is an acceptable match for the planet's radius with NEA values. Also, based on the average value of 350 mm optics in this study, it shows that the results obtained using small telescopes can be very significant if there is appropriate observational skill to study more discovered planets.

**Introduction**
Throughout the transit of an exoplanet passing in front of the disk of its' host star, it blocks a small portion of light from the star with the change in flux at mid-transit known as the transit depth (Rice, 2014). Therefore, photometric observation of transit and the data of the transit light curve can then be used to deduce the orbital and physical parameters of the exoplanets. Ground-based observations have great scientific potential in the discovery of new exoplanets and also have the follow-up potential in providing system parameters as well as mid-transit times (Davoudi, 2020).

It is possible to get ground-based observation raw data from Exoplanet Transit Database (ETD)[1] that is preserved by the Czech Astronomical Society (CAS). The ETD has been established to collect all available transit data with different qualities which then classify them. This online portal provides some useful information for observers such as time of the transit start, center, end, duration, and the depth of each transit that named as transit prediction. Moreover, it has an algorithm for processing photometric data which fits them to a light curve. The planetary transit measured in the light curve is mostly described by three parameters: depth, duration, and its' mid-transit time (Poddaný, 2010).

Throughout this study, we investigated 10 exoplanets and their light curves in proportion to the ETD. The exoplanets that we chose to discuss in this paper are alone in their planetary system or have distant companions. They also all belong to hot Jupiter type planets with an apparent magnitude between 9.8 and 14.07. We have studied Corot-12 b (Gillon et al., 2010), HAT-P-52 b (Hartman et al., 2015), HAT-P-57 b (Hartman et al., 2015), HATS-28 b (Espinoza et al., 2016), HATS-34 (de Val-Borro et al., 2016), KELT-3 b (Pepper et al., 2013), WASP-61 b (Hellier et al., 2012), WASP-67 b (Hellier et al., 2012), WASP-122 b (Rodriguez et al., 2016) and WASP-140 b (Hellier et al., 2016) whose their detection are primary transit and their light curves are taken from ETD.

We know hot Jupiters as a class of giant gassy exoplanets. They are the easiest planets to detect because of their large size (with masses greater than or equal to 0.25 Jupiter mass) and short orbital period (with period between 0.8 - 6.3 days (Becker et al., 2015). There are several mechanisms to explain some of the observational properties of hot Jupiters. The main properties can be summarized as follows: a) orbital periods about three days, b) variety of obliquity, c) solitary host stars, d) distinct mass functions from other types of exoplanets, e) narrow range of stellar mass. For more information about the hot Jupiters, you can refer to the following paper (Martin et al., 2011).

---
[1]http://var2.astro.cz/ETD/

The ETD adopts the parameter uncertainties from a Levenberg-Marquardt optimization algorithm (Poddaný et al., 2010), which is believed to be unreliable in the presence of parameter correlations (Mallonn et al., 2019). So we reanalyzed ETD light curves using an exact method. The activity described in this study is done in two sections. In the first part, we chose some exoplanets and then analyzed and reduced observational data from their light curves. In the next section, we compared values of the output parameters of planets from a web-based tool (A Fast Exoplanetary Fitting Suite in IDL) with the parameters from the Extrasolar Planets Encyclopedia[2] database. EXOFAST[3] as a fast transit parameter fitter takes flux and time in $BJD_{TDB}$ (Barycentric Julian Dates in Barycentric Dynamical *Time*) as inputs and provides Stellar, planetary, and primary transit parameters.

EXOFAST is an IDL library for transit and radial velocity modeling and portrays the parameter uncertainties with a differential evolution Markov Chain Monte Carlo method (MCMC), which is presented by Eastman, Gaudi, and Agol (2013). The authors acquire noteworthy improvement in the evolution speed of the quadratic MA model by swapping the modern way including the calculation of elliptic of integral of the third kind with a faster one. The transit method has been highly significant for this revolution. It can be utilized to find out orbital inclination, planet radius, mass, and average density of the planet (Guillot, 2005; Sato et al., 2005; Charbonneau et al., 2006; Fortney et al., 2006). Moreover, we can obtain some information about their atmospheres (Charbonneau et al., 2002; Vidal-Madjar et al., 2003), thermal emission (Deming et al. 2005; Charbonneau et al., 2006), and also provide a probe of exoplanetary structures (Ragozzine & Wolf, 2009; Carter & Winn, 2010).

**Observation**

The first step to derive the orbital and physical parameters of exoplanets is observation. In this project, we collected light curves of ground-based observations that are available on the ETD website. So we found the raw dataset of the exoplanets and then data reduction was done.

In Table 1, we present the characteristics of the host stars and the planets, which is based on the Extrasolar Planets Encyclopaedia. We put the observation's information in Table 2 which includes the planet's name, the planet's celestial coordinates (Right Ascension (RA.) and Declination (Dec.)), the observer's name and the date on which the mid-transit happened. Also, we put some information about the observation tools the observers used to collect the exoplanet's transit data; they include optical telescope which we wrote its' optic size, Charged Couple Device (CCD) model and also the filter which is used up to the observer's choice.

**Table 1. Specifications of the host stars and the planets. All of these exoplanets detected by Primary Transit method.**

| Star Name | $RA_{2000}$ | $Dec_{2000}$ | D. (pc) | Spect. | App. Mag. | Planet | Discovered |
|---|---|---|---|---|---|---|---|
| CoRoT-12 | 06 43 03.76 | -01 17 47.14 | 1150 | G2V | 15.52 | CoRoT-12 b | 2010 |
| HAT-P-52 | 02 50 53.20 | +29 01 20.52 | 385 | - | 14.07 | HAT-P-52 b | 2015 |
| HAT-P-57 | 18 18 58.42 | +10 35 50.12 | 303 | - | 10.47 | HAT-P-57 b | 2015 |
| HATS-28 | 18 57 35.92 | -49 08 18.55 | 521 | G | - | HATS-28 b | 2016 |
| HATS-34 | 00 03 05.87 | -62 28 09.61 | 532 | - | 13. 85 | HATS-34 b | 2016 |
| KELT-3 | 09 54 34.38 | +40 23 16.97 | 178 | F | 9.8 | KELT-3 b | 2012 |
| WASP-61 | 05 01 11.91 | -26 03 14.96 | 480 | F7 | 12.5 | WASP-61 b | 2011 |
| WASP-67 | 19 42 58.52 | -19 56 58.52 | 225 | K0V | 12.5 | WASP-67 b | 2011 |
| WASP-122 | 07 13 12.35 | -42 24 35.11 | 251.93 | G4 | 11.0 | WASP-122 b | 2015 |
| WASP-140 | 04 01 32.54 | -20 27 03.91 | 180 | K0 | 11.1 | WASP-140 b | 2016 |

---

[2] http://exoplanet.eu/

[3] https://exoplanetarchive.ipac.caltech.edu/cgi-bin/ExoFAST/nph-exofast

Table 2. Initial data information for planets.

| Planet | RA$_{2000}$ - DEC$_{2000}$ | Observer | Observation date | Filter | Optic size (mm) | CCD |
|---|---|---|---|---|---|---|
| CoRot-12 b | 06:43:03.76 -01:17:47.14 | F. Grau Horta | 2014-12-23 | R | 305 | FLI PL1001E-1 |
| HAT-P-52 b | 02:50:53.20 29:01:20.52 | F. Campos | 2018-01-21 | Clear | 350 | ST-8XME |
| HAT-P-52 b | 02:50:53.20 29:01:20.52 | R. Naves | 2016-12-02 | Clear | 305 | Moravian G4 |
| HAT-P-52 b | 02:50:53.20 29:01:20.52 | Y. Jongen | 2019-01-03 | Clear | 425 | STLX11002 |
| HAT-P-57 b | 18:18:58.43 10:35:50.13 | A. Wünsche | 2019-06-29 | V | 820 | FLI PL230 |
| HAT-P-57 b | 18:18:58.43 10:35:50.13 | F. Lomoz | 2019-07-03 | V | 300 | ST2000XM |
| HATS-28 b | 18:57:35.93 -49:08:18.56 | P. Evans | 2016-07-02 | Clear | 250 | ST9XE |
| HATS-34 b | 00:03:05.87 -62:28:09.62 | Y. Jongen | 2020-01-23 | Clear | 425 | Moravian 4G |
| KELT-3 b | 09:54:34.38 40:23:16.98 | A. Ayiomamitis | 2013-03-09 | Clear | 305 | ST-10XME |
| KELT-3 b | 09:54:34.38 40:23:16.98 | S. Gudmundsson | 2014-02-27 | Clear | 300 | STL-11k |
| WASP-61 b | 05:01:11.92 -26:03:14.97 | C. Quiñones, et al. | 2014-10-19 | B | 1540 | - |
| WASP-67 b | 19:42:58.52 -19:56:58.52 | P. Evans | 2014-05-20 | Clear | 250 | ST9XE |
| WASP-122 b | 07:13:12.35 -42:24:35.12 | Y. Jongen | 2020-01-14 | V | 425 | Moravian 4G |
| WASP-140 b | 04:01:32.55 -20:27:03.91 | F. Lomoz | 2017-01-01 | Clear | 254 | G2-8300 |

**Method and analysis**

We used the raw data for 10 exoplanets that consist of three columns of time in Julian Dates (JD) or Heliocentric Julian Dates (HJD), Delta Magnitude, and its' error. JD or HJD time frame was converted to BJD _TDB through Time Utilities[4], applying RA, DEC of the host star from the Simbad website[5]. Then we used Phoebe software to turn Delta Magnitude to normalized flux.

We prepared a file including time in BJD$_{TDB}$, Flux, and its' Error and then used it as an input for EXOFAST. After importing the data file to EXOFAST, the filter type of each observation was selected. Then, we disabled the Include all Detrending Columns section. Three parameters of metallicity, effective temperature, and Log g are necessary for EXOFAST. EXOFAST has done the best fit on light curves (Figure 1. and Figure.2) and calculated the parameters of transit based on the given data. The whole light curves were displayed in the Appendix. The horizontal axis is Time (BJD$_{TDB}$)-T$_c$ based on hours. The vertical axis in the top panel is based on the normalized flux of the star and the O-C which means the residuals between observed and fitted data points.

The output parameters of EXOFAST and their uncertainties were listed in Table 3. The second column represents the planet's radius R$_p$ (R$_J$), the third column contains the ratio of the planet radius to the stellar radius $k = \frac{R_p}{R_*}$, the fourth column is transit center time T$_C$ (BJD$_{TDB}$), the fifth column is FWHM duration (days) T$_{FWHM}$, transit depth δ, and $\tau$, Ingress/egress duration are shown in the sixth and seventh columns respectively. We calculated the uncertainties of k and R$_p$ from the following equations:

$$\sigma_k = \frac{\sigma_{depth}}{2k}, \qquad \sigma_{R_p} = k\sigma_{R_*} + R_*\sigma_k \qquad (1)$$

where $R_*$ and its uncertainty extracted from NEA[6] in Table 3.

---

[4]http://astroutils.astronomy.ohio-state.edu/time/
[5]http://simbad.u-strasbg.fr/simbad/
[6]https://exoplanetarchive.ipac.caltech.edu/

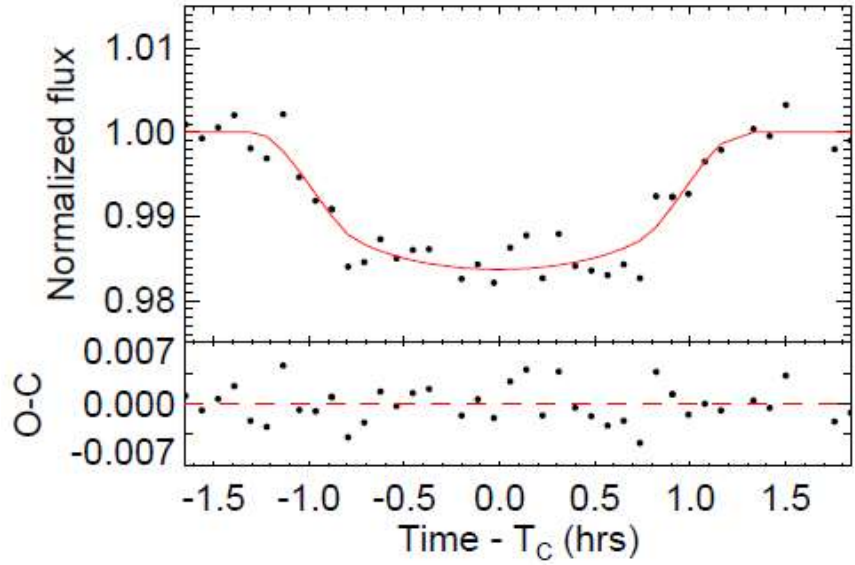

**Figure 1.** The observational and theatrical light curves of HAT-P-52 b extracted by EXOFAST software.

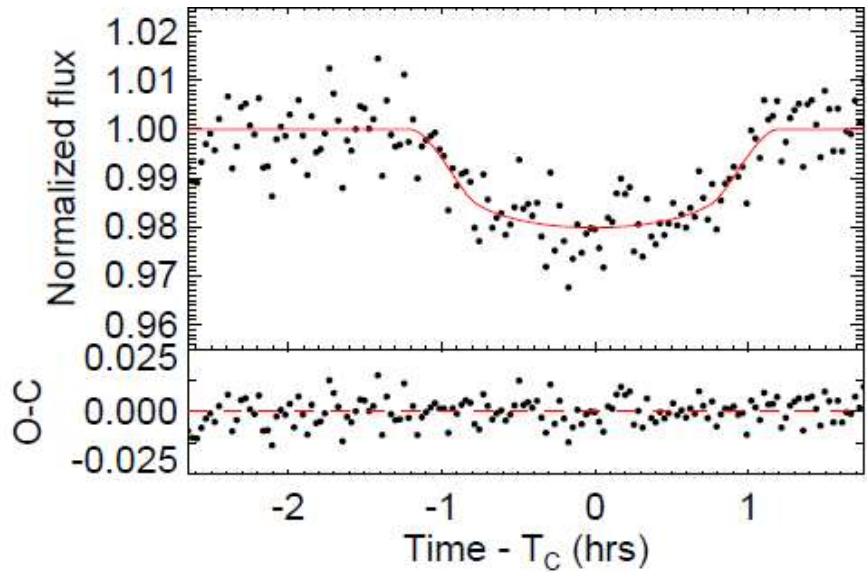

**Figure 2.** The observational and theatrical light curves of HAT-P-28 b extracted by EXOFAST software.

**Table 3.** Exofast's results for 10 hot Jupiter exoplanets in this study.

| Planets | $R_P(R_J)$ | K | $T_C(BJD_{TDB})$ | $T_{FWHM}$ | $\delta$ | $\tau$ |
|---|---|---|---|---|---|---|
| CoRot-12 b | 1.8058 (±0.2109) | 0.1685 (±0.0055) | 2457014.58 (±0.0022) | 0.1055 (±0.0045) | 0.0284 (±0.0018) | 0.0223 (±0.0078) |
| HAT-P-52 b | 1.0964 (±0.0750) | 0.1241 (±0.0021) | 2458140.343 (±0.0010) | 0.0852 (±0.0020) | 0.0154 (±0.0005) | 0.0206 (±0.0034) |
| HAT-P-52 b | 1.1418 (±0.0874) | 0.1293 (±0.0032) | 2457724.553 (±0.0012) | 0.0950 (±0.0024) | 0.0167 (±0.0008) | 0.0123 (±0.0042) |
| HAT-P-52 b | 0.8778 (±0.0687) | 0.0994 (±0.0026) | 2458520.336 (±0.0009) | 0.0754 (±0.0018) | 0.0098 (±0.0005) | 0.0081 (±0.0032) |
| HAT-P-57 b | 1.2333 (±0.0602) | 0.0937 (±0.0010) | 2458663.509 (±0.0006) | 0.1398 (±0.0013) | 0.0087 (±0.0002) | 0.0132 (±0.0022) |
| HAT-P-57 b | 1.4751 (±0.1124) | 0.1122 (±0.0039) | 2458668.44 (±0.0023) | 0.1353 (±0.0045) | 0.0126 (±0.0009) | 0.0153 (±0.0079) |
| HATS-28 b | 1.2139 (±0.0750) | 0.1337 (±0.0025) | 2457571.884 (±0.0010) | 0.0822 (±0.0020) | 0.0179 (±0.0006) | 0.0171 (±0.0035) |
| HATS-34 b | 1.3289 (±0.0950) | 0.1364 (±0.0034) | 2458871.602 (±0.0013) | 0.0315 (±0.0026) | 0.0186 (±0.0009) | 0.0315 (±0.0046) |
| KELT-3 b | 1.4832 (±0.0747) | 0.1023 (±0.0007) | 2457126.464 (±0.0005) | 0.1157 (±0.0011) | 0.0104 (±0.0001) | 0.0235 (±0.0019) |
| KELT-3 b | 1.3706 (±0.1032) | 0.0945 (±0.0030) | 2456715.552 (±0.0019) | 0.1092 (±0.0038) | 0.0089 (±0.0005) | 0.0155 (±0.0066) |
| WASP-61 b | 1.1478 (±0.0474) | 0.0873 (±0.0016) | 2456950.747 (±0.0013) | 0.1612 (±0.0025) | 0.0076 (±0.0003) | 0.0141 (±0.0044) |
| WASP-67 b | 0.9970 (±0.0676) | 0.0986 (±0.0034) | 2456798.021 (±0.0009) | 0.0619 (±0.0019) | 0.0097 (±0.0007) | 0.0061 (±0.0033) |
| WASP-122 b | 1.7442 (±0.0802) | 0.1208 (±0.0013) | 2458862.643 (±0.0007) | 0.0508 (±0.0013) | 0.0146 (±0.0003) | 0.0390 (±0.0024) |
| WASP-140 b | 1.1990 (±0.0735) | 0.1406 (±0.0022) | 2457755.319 (±0.0006) | 0.0423 (±0.0012) | 0.0197 (±0.0006) | 0.0175 (±0.0021) |

Table 4. The radius of the host stars from NEA.

| Planets | $R_P(R_J)$ |
|---|---|
| CoRot-12 b | 1.44±0.13 (Gillon et al., 2010) |
| HAT-P-52 b | 1.009±0.072 (Hartman et al., 2015) |
| HAT-P-57 b | 1.413±0.054 (Hartman et al., 2015) |
| HATS-28 b | 1.194±0.07 (Buchhave et al., 2011) |
| HATS-34 b | 1.43±0.19 (de Val-Borro et al., 2016) |
| KELT-3 b | 1.345±0.072 (Pepper et al., 2013) |
| WASP-61 b | 1.24±0.03 (Hellier et al., 2012) |
| WASP-67 b | $1.4^{+0.3}_{-0.2}$ (Hellier et al., 2012) |
| WASP-122 b | 1.743 ± 0.047 (Turner et al., 2016) |
| WASP-140 b | $1.44^{+0.42}_{-0.18}$ (Hellier et al., 2016) |

The positions of all host stars in this study, in which the theoretical Zero Age Main Sequence (ZAMS) and Terminal Age Main Sequence (TAMS), are shown in the H-R diagram in Figure 3 and the calculation are shown in Table 5. As can be seen in the diagram, most of the stars in this study are in the middle or second part of their lifetime, and the planets were discovered by primary transit at the time when star's temperature is cooler than the first part of their lifetime.

Table 5. Calculations of host stars and parameters determining their position in the H-R diagram.

| Planet Name | T (K) | R ($R_{Sun}$) | L | Log $T_{eff}$ | Log ($L/L_{sun}$) |
|---|---|---|---|---|---|
| CoRoT-12 b | 5675.0 ± 80 (Gillon et al., 2010) | | 1.159 ± | 3.7539 | 0.0640 |
| HAT-P-52 b | 5131.0 ± 50 ( Hartman et al., 2015) | | 0.4959 ± | 3.7102 | -0.3046 |
| HAT-P-57 b | 6330 ±124 (Stassun et al., 2017) | | 3.241 ± | 3.8014 | 0.5106 |
| HATS-28 b | 5498.0 ± 84 (Espinoza et al., 2016) | | 0.6969 ± | 3.7402 | -0.1568 |
| HATS-34 b | 5380.0 ± 73 (de Val-Borro et al., 2016) | | 0.7219 ± | 3.7307 | -0.1415 |
| KELT-3 b | 6304.0 ± 49 (Stassun et al., 2017) | | 3.112 ± | 3.7996 | 0.4930 |
| WASP-61 b | 6250.0 ± 15 (Stassun et al., 2017) | | 2.532 ± | 3.7958 | 0.4034 |
| WASP-67 b | 5200.0 ± 10 (Stassun et al., 2017) | | 0.4965 ± | 3.7160 | -0.3040 |
| WASP-122 b | 5720 ±13 (Turner et al., 2016) | | 2.3584 ± | 3.7573 | 0.3726 |
| WASP-140 b | 5260.0 ± 10 (Hellier et al., 2017) | | 0.5198 ± | 3.7209 | -0.2841 |

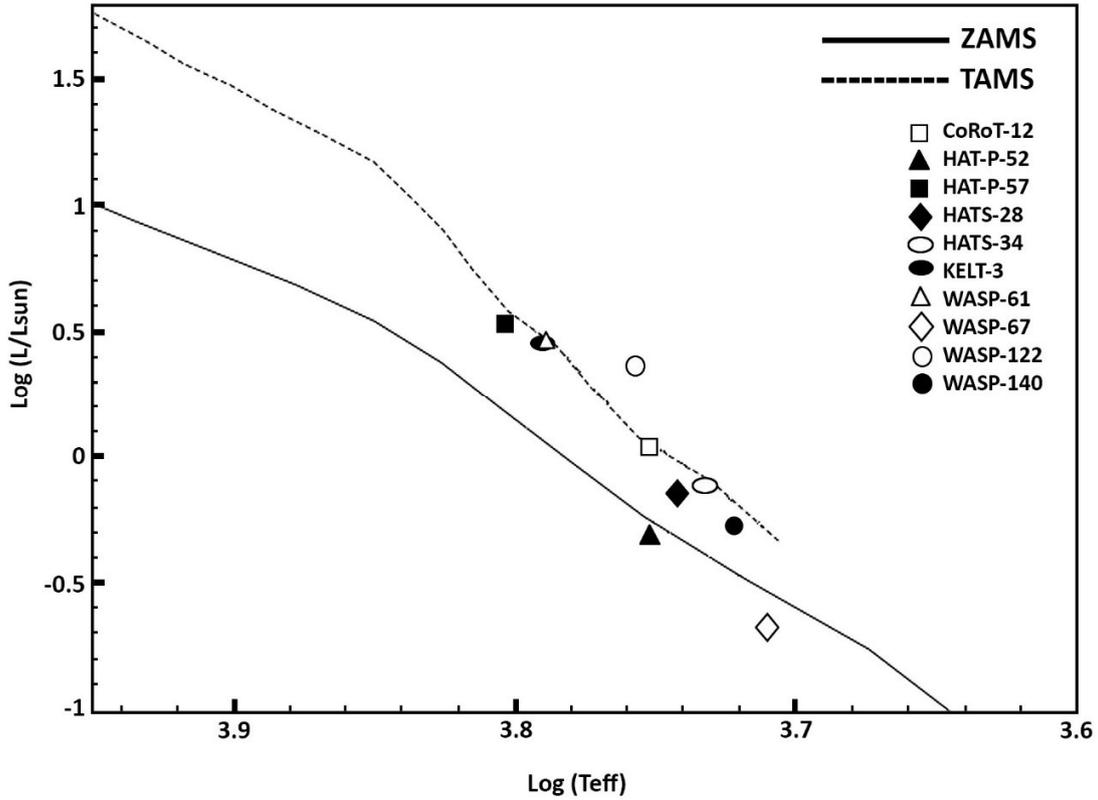
Figure 3. The position of the host stars on the H-R diagram.

## Conclusion

In order to compare the parameters of exoplanets obtained from the analysis of data related to ground observations with small telescopes to the parameters available in the NASA archive (NEA), we analyzed 14 transit light curves for 10 hot Jupiter exoplanets from the ETD database.

The telescopes used in the observations have an average aperture of 350 mm. The observations were used CCD method of observation. After the raw data reduction, we prepared a file including time in $BJD_{TDB}$, Flux, and its' Error and then used it as an input for EXOFAST and extracted the parameters of exoplanets. The radius of the exoplanet is the most important parameter determined by the transit method. Using the confidence interval method, the radius of exoplanet has a good agreement with NEA in $1\sigma$ to $3\sigma$.

The results show that the radius of exoplanets obtained from ground-based observation with small telescopes are comparable to its value in the NEA in this study. This is important because it can show the role of observations with small telescopes to study more discovered planets.

## Acknowledgments

This manuscript was prepared by the International Occultation Timing Association Middle East section (IOTA/ME) and with participants at a project was held as a scientific activity on exoplanets during winter and spring of 2020.

## Appendix

In the figures, the light curves of planets were displayed in different bands extracted by EXOFAST software. The horizontal axis is Time ($BJD_{TDB}$)-$T_c$ based on hours. The vertical axis in the top panel is based on Normalized flux of the star and the O-C which means the Residuals between observed and fitted data points.

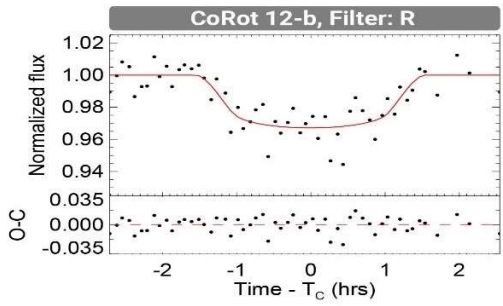
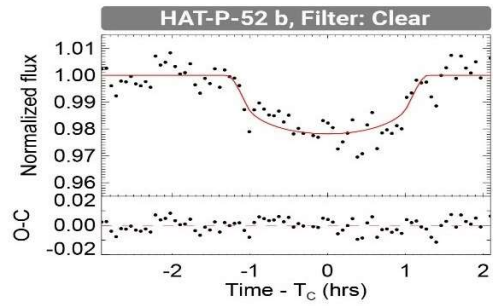
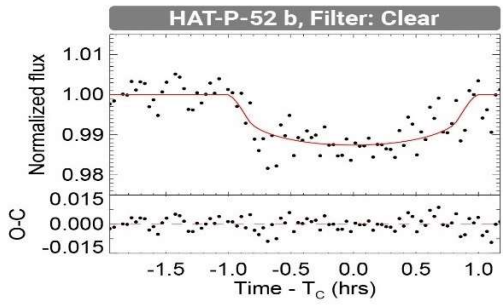
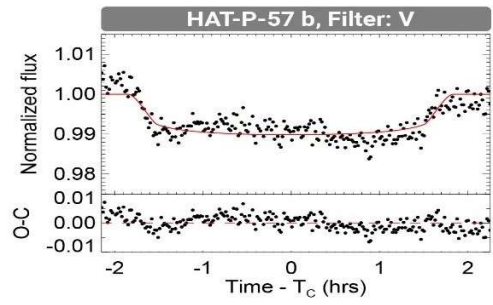
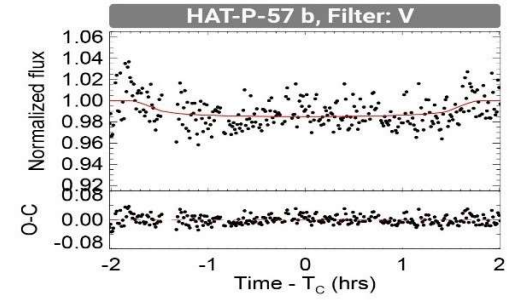
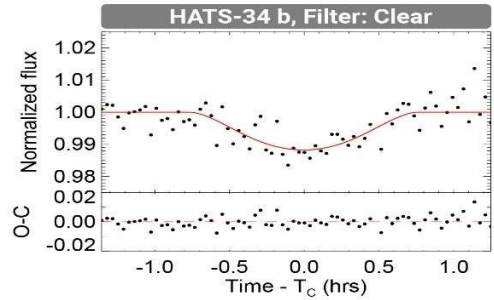
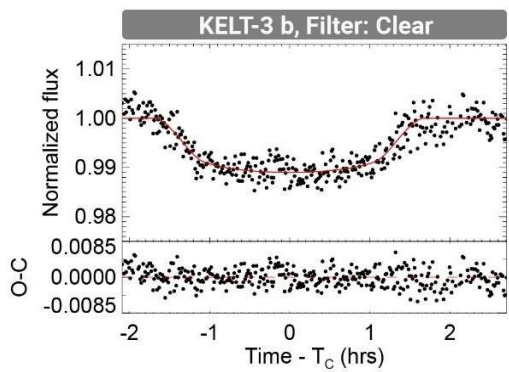
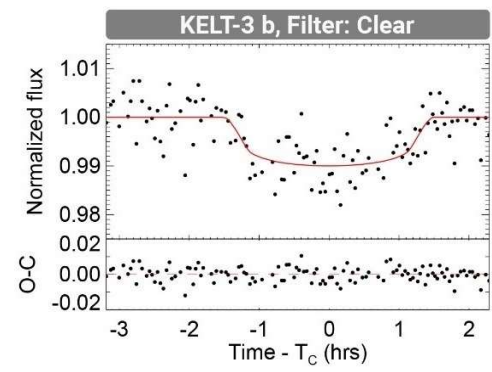

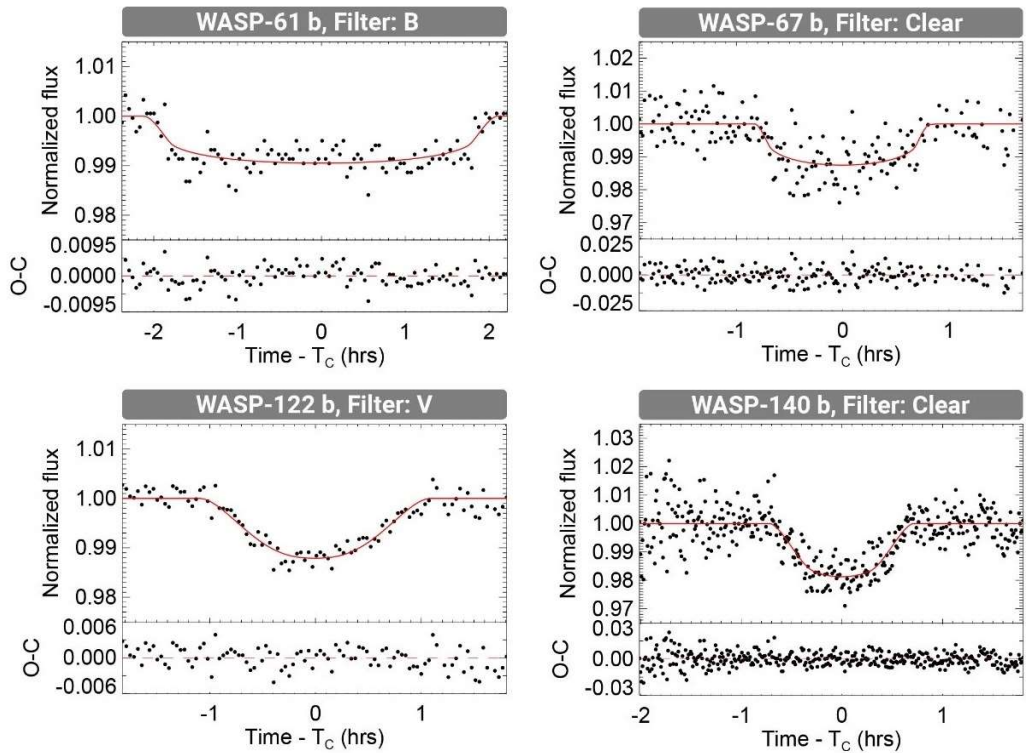